\documentclass[conference]{IEEEtran}
\IEEEoverridecommandlockouts
\usepackage{amsmath,amssymb,amsfonts}
\usepackage{algorithmic}
\usepackage{graphicx}
\usepackage{textcomp}
\usepackage{enumitem}
\usepackage{xcolor}
\usepackage[sorting=ynt]{biblatex}
\def\BibTeX{{\rm B\kern-.05em{\sc i\kern-.025em b}\kern-.08em
    T\kern-.1667em\lower.7ex\hbox{E}\kern-.125emX}}

\begin{document}

\title{Quantum Annealing for Vehicle Routing Problem with weighted Segment \\

}

\author{\IEEEauthorblockN{ Toufan Diansyah Tambunan\textsuperscript{(1)}, Andriyan Bayu Suksmono, Ian Joseph Matheus Edward, Rahmat Mulyawan}
\IEEEauthorblockA{\textit{School of Electrical Engineering and Informatics} \\
\textit{Institut Teknologi Bandung (ITB).}\\
 Bandung, Indonesia \\
\textsuperscript{(1)}tambunan@tass.telkomuniversity.ac.id}

}

\maketitle

\begin{abstract}
Quantum annealing technologies aim to solve computational optimization and sampling problems. QPU (Quantum Processing Unit) machines such as the D-Wave system use the QUBO (Quadratic Unconstrained Binary Optimization) formula to define model optimization problems for quantum annealing. This machine uses quantum effects to speed up computing time better than classical computers. We propose a vehicle routing problem that can be formulated in the QUBO model as a combinatorial problem, which gives the possible route solutions increases exponentially. The solution aims to optimize the vehicle's journey to reach a destination. The study presents a QUBO formulation to solve traffic congestion problems on certain roads. The resulting route selection by optimizing the distribution of the flow of alternative road vehicles based on the weighting of road segments. Constraints formulated as a condition for the level of road density. The road weight parameter influences the cost function for each road choice. The simulations on the D-Wave quantum annealer show optimal results on the route deployment of several vehicles. So that each vehicle will be able to go through different road options and reduce road congestion accurately. This solution provides an opportunity to develop QUBO modeling for more complex vehicle routing problems for road congestion.
\end{abstract}


\section{Introduction}
The development of quantum annealing is designed to solve combinatorial optimization problems \cite{kadowaki1998quantum}, which allows finding solutions in a way that is better than classical computers. The quantum annealing technology was first developed for public and commercial by D-Wave Systems. This system exploits the advantages of quantum mechanisms such as entanglement, superposition and implements quantum bits (qubits) for the computational process in sampling and optimization problems \cite{farhi2000quantum}. The existence of quantum effects inspires the annealing process to avoid local minima of the cost function by tunneling the effect through the barrier that separates local minima (William, 2019). The optimization approach is based on the observation that the cost function of the optimization problem can be seen as the energy of the physical system, and the energy barrier can be traversed by thermal hopping, also known as tunneling. The optimization required to avoid the local minima trap can be an advantage of exploring the low energy configuration of quantum mechanics using superposition and tunneling. The quantum annealing mechanism works on this idea and has been introduced as an algorithm that can solve optimization problems such as finding the ground states of a spin Hamiltonian \cite{kadowaki1998quantum}. Quantum computers (such as D-Wave) have proven to be an advantage for solving various NP-hard problems. The way that a quantum annealer tries to solve problems is very similar to how optimization problems are solved using (classical) simulated annealing \cite{kadowaki2002study}. The energy landscape is constructed through multivariate functions so that the ground state corresponds to the solution of the problem.

The quantum annealer machine can solve optimization problems by modeling expressed in the form of QUBO (quadratic unconstrained binary optimization) \cite{boros2007local}. This model belongs to a pattern matching technique that can be used in machine learning and optimization to minimize the quadratic polynomial over binary \(\{0,1\}\) variables \cite{papalitsas2019qubo}. The QUBO model is similar to the Ising model in that the underlying schema of the model can be depicted in a graph with qubits as vertices and couplers as edges connecting qubits. The variables representing the qubits and the interactions between the qubits represent the costs associated with each pair of qubits. Quantum annealing research related to optimization has been carried out, including in graph partition problems \cite{ushijima2017graph}, jobs scheduling optimization \cite{venturelli2016job}, application of solutions for max clique cases on graphs \cite{chapuis2019finding}, finding Hadamard matrices with QAM \cite{suksmono2019finding}, optimization for the traveling-salesman problem \cite{martovnak2004quantum},regulation of road density on traffic flow problems \cite{neukart2017traffic}, road traffic research \cite{hussain2020optimal} and \cite{inoue2021traffic} by optimizing traffic lights to reduce road congestion.

This paper presents a QUBO formulation for traffic flow problems to find alternative routes for vehicles on the road. This problem is part of the route finding solution on vehicles that requires a combinatorial solution. Here we show that while our current work is an extension of the previous study by Neukart et al., we approach it differently. In the previous work \cite{neukart2017traffic,horvat2015traffic}, minimizing traffic congestion was achieved by assigning the most optimal route to each vehicle to reduce overlapping routes in each vehicle itinerary with those in other vehicle routes. In our traffic flow model, the vehicle's route is affected by the weight assigned to each road segment. Congestion on road conditions is influenced by the number of vehicles, with the following definitions:
\begin{enumerate}
    \item The condition of the number of vehicles sharing the lane segment at a certain time.
    \item Minimization of the number of overlapping segments on the route of each vehicle.
    \item A cost function is obtained from the number of vehicles sharing lanes in one lane segment.
\end{enumerate}
The cost value will affect the density level of the selected road segment to avoid traffic jams.

This model can then be used as a property of the road conditions in real terms. The property of road conditions can be the assumed distance between points, or the density level of the segment, or other parameters that determine segment selection. This problem is often encountered in the case of traffic management in several vehicles heading to the same endpoint and potentially congestion at the intersection.

To solve a problem using the D-Wave quantum annealer, we must define it as a QUBO problem or an equivalent Ising function defined on logical variables. We embed the logical problem in the physical architecture of the quantum annealer by mapping logical variables and qubits. The final step consists of performing an annealing process and obtaining the results. The D-Wave quantum annealer has some limitations, such as the maximum number of physical qubits available (with 2.048 qubits in its most recent computer, the D-Wave 2000Q system). This limitation imposes a restriction on the size of the logical problem that can be embedded into the quantum annealer machine.

\section{Methods}
\subsection{Quantum Annealing with D-Wave Systems}
QA (quantum annealing) is implemented in hardware to manipulate a collection of quantum bit (qubit) states according to a time-dependent Hamiltonian Ising model. In the workings of D-Wave systems \cite{lucas2014ising}, the annealing mechanism works using the Hamiltonian Ising model concept, which represents the following notation.

\begin{equation}
\small H = - \frac {A(s)} {2} {\left ( \sum _{i}   {\widehat{\sigma}}_{x} ^{(i)}   \right )} + \frac {B(s)} {2} {\left ( \sum _{i}   h_i {\widehat{\sigma}}_{z} ^{(i)} +   \sum _{i>j}   J_i,j {\widehat{\sigma}}_{z} ^{(i)} {\widehat{\sigma}}_{z} ^{(j)} \right )}
\label{eq:hamilton}
\end{equation}
Where \(\sigma \) is the pauli matrix for the \(q\) qubit operations, while \(h\) and \(j\) are the qubit bias and coupling strengths. Hamiltonian notation \cite{headquarters2021d} consists of the initial Hamiltonian equation (equation in the first part) to the final Hamiltonian (at the end of the equation). The initial Hamiltonian is the system's initial state with the lowest energy. In this initial state, all qubits are in the superposition state 0 and 1. The initial Hamiltonian is also known as the tunneling Hamiltonian, while the final Hamiltonian is a classic condition of the problem. Then the system will go to the final Hamiltonian condition, where the lowest-energy position in the final Hamiltonian is the solution to the problem being solved. The final state, in this case, is the condition of the classical system, which is influenced by qubit biases and coupling strengths. This situation is also known as the Hamiltonian Ising problem. The optimal solution is found in the annealing process in the Hamiltonian problem. Finding solutions for QA on the D-Wave system \cite{headquarters2021d} is done by creating Binary Quadratic Models (BQM), then mapping the problem modeling in the form of Ising model or quadratic unconstrained binary optimization (QUBO). The Ising model or QUBO form is used to formulate BQM, consisting of objective and constraint function components. QUBO maps the problem in the form of a boolean variable so that it can maximize the objective function in a linear expression \cite{lewis2017quadratic}. QUBO formula for problem optimization:
\begin{equation}
Obj( x, Q ) = x^T \cdot Q \cdot x 
\label{eq:qubo}
\end{equation}
where \(x\) is a vector of binary variables of size \(N\), and \(Q\) is an \(N\times N\) (upper-triangular) matrix that expresses the relationship between variables. The optimization search on the objective function is the same as the minimal function on the Ising model problem \cite{lucas2014ising}. The \(x\) values in the QUBO  and Ising models  represent different ways. The following optimization function defines the matrix Q in the upper-triangular form\cite{glover2018tutorial}:
\begin{equation}
f ( x_1, ... , x_n ) = \sum _{i=1}^{n} Q_{i,i} {x_i} + \sum _{1 \leq i < j \leq n} Q_{i,j} {x_i}{x_j}
\label{eq:matrixQ}
\end{equation}
The diagonal representation of the matrix \(Q_{i,i}\)is a linear coefficient value, while the top (non-zero) of the \(Q\) matrix is a quadratic coefficient. Furthermore, from the resulting QUBO Matrix, the binary variable value will be searched, which in combination can give the most minimal objective function equation results. This model is adopted on the workings of quantum annealing with the application of qubits, couplers, and entanglements on the D-Wave systems. The objective function in D-Wave BQM is a problem to be optimized (minimum cost). Meanwhile, the function constraints are limitations that the solution must meet according to the rules of the problem.

\subsection{QUBO Formulation for Routing Optimization}
The objective of the traffic flow for vehicle routing optimization problem is to minimize the time for a given set of cars to travel between their individual sources and destinations. We used the simplifying assumption that the time to traverse a street is proportional to a function of the number of cars currently occupying the street. Thus, we minimize the total time for all cars by minimizing overall road segments' congestion. Congestion on an individual road segment is determined by a quadratic function of the number of cars traversing it in a specific road segment. To optimize the traffic flow, we minimize the number of overlapping road segments between assigned routes for each car. We formulate an optimization problem by giving three possible routes for each car and a choice of car routes that minimize congestion on all road segments. We require that each car must be assigned one of three route options. At the same time, this option can minimize the total congestion on all specified routes. It is important to emphasize that each car is proposed three alternative routes in this example, which may not be the same set of 3 routes for all cars.
\begin{figure}[htbp]
\centering
\includegraphics[width=0.48\textwidth, keepaspectratio]{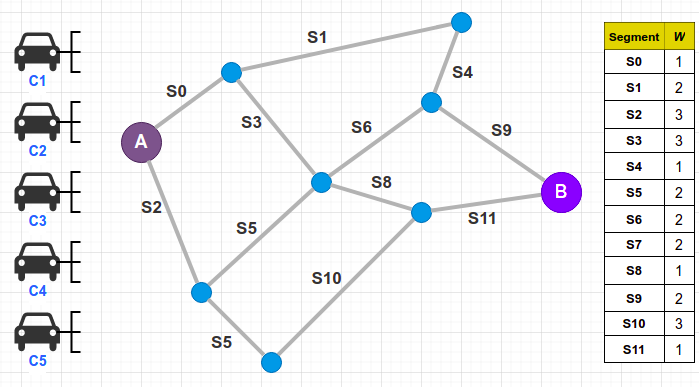}
\caption{illustration of vehicle route optimization problem from point A to B}
\label{fig 1route_problem}
\end{figure}

The road conditions in Fig.1 show vehicle traffic from point A to point B, consisting of 9 nodes and 12 edges. Each vehicle is given three choices of alternative routes, but only one route is chosen by a vehicle as a road solution that can be passed. The choice of route will have a minimum of 4 segments to pass from point A to point B. Vehicles are represented by variable \(i=\{1,2,3, ..., n\}\), and route choices are given for each vehicle with three choices of route \(j=\{1, 2, 3\}\). The combination of each route choice from the vehicle will be modeled in the form of a variable \(q_{i,j}\) which represents qubits. Each vehicle will pass through a selection of segments \(S=\{s_1, s_2, s_3, ..., s_m\}\), according to the route chosen from point A to reach the destination at point B. In this case, weight is added to the segment, which has a value of \(s_m \geq 1\). 
\begin{table}[htbp]
\caption{Combination of qubits (Q) of route choices (j) vehicles (i) that will traverse the segment (S).}
\begin{center}
\begin{tabular}{c}
\includegraphics[width=0.45\textwidth, keepaspectratio]{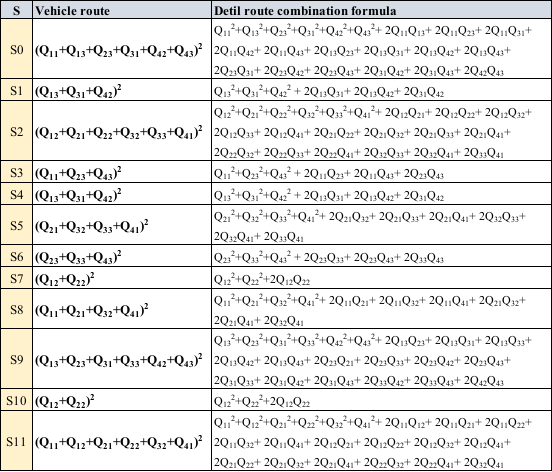}
\end{tabular}
\label{tab2}
\end{center}
\end{table}

The weights on the segments have different values and can be assumed as parameters of distance, road priority, or other importance. If the road segment is selected, it will be worth 1 (true); while not selected, it will be worth 0 (false).
\begin{equation}
w_{ij} = \sum _{m=0}^{11} s_m , {q_{ij}(m)=1}
\label{eq:weight_segmen}
\end{equation}
The accumulated segment \(s_m\) that are selected on the vehicle route \(q_{ij}\) will be weighted \(w_{ij}\), which means the sum of the weights of each segment passed by the vehicle. The vehicle's weight on the choice of road route \(w_{ij}\) will affect the use of the QUBO matrix in the cost function of this case. The condition that must be met from this problem is that each vehicle must choose exactly one route, which will be used as a constraint component in the QUBO model. Simulations that refer to previous research problems (Neukart, 2019), the use of \(n\) = 4 vehicles, it takes 12 qubits of variables to be able to represent the flow of vehicles (each vehicle has \(j\)=3 alternative routes). Each segment (\(s_m\)) can be crossed by several possible vehicle routes (\(q_{ij}\)) and will be accumulated to calculate the density level of the segment \(s_m\). The following table describes the combination of route options for each vehicle  (qubit \(Q\)) that traverses the road segment \(s_m\).

The development of this research is by adding a weight parameter (\(w_{ij}\)) on each road segment which can be used as a parameter of distance, road priority, or other interests. The weighting of the road segment will affect the cost calculation. From what was previously only influenced by the density of vehicles in a segment (\(s_m\)), the weight (\(w_{ij}\)) of the road segment will also affect it. The following equation is the cost function for the road segment traversed by the vehicle.
\begin{equation}
cost(s_m) = \left ( \sum _{q_{ij} \in B_{s_m}} {w_{ij} q_{ij}} \right )^2
\label{eq:cost_func}
\end{equation}

The configuration of the route choice for each vehicle can form an overall cost function equation by accumulating the weight \(w_{ij}\) on each segment which is the route of choice for the vehicle \(q_{ij}\). The number of segments that overlap with other vehicles is minimized to optimize vehicle traffic. So, three alternative routes are proposed for each vehicle different from other vehicles. The cost function of each segment \(cost(s_m)\) is calculated from the number of vehicles using that segment. The cost value will affect the density level of the selected road segment to avoid congestion. Alternative vehicle routes are selected based on the least cost value.
\begin{equation}
1 = \left ( \sum _{j=1}^{3} {q_{ij}} \right )^2 
\label{eq:constraint_func1}
\end{equation}\begin{equation}
0 = \left ( \sum _{j=1}^{3} {q_{ij} - 1} \right )^2 
\label{eq:constraint_func2}
\end{equation}\begin{equation}
0 = ( -q_{i1}-q_{i2}-q_{i3} + 2 q_{i1}q_{i2} + 2 q_{i1}q_{i3} + 2 q_{i2}q_{i3} + 1) 
\label{eq:constraint_func3}
\end{equation}

The constraint function in the above equation will get 0 when the condition is true when only one vehicle route option is active (\(q_{ij}\) is 1). The problem of vehicle traffic flow is defined by a condition rule where each vehicle (\(i\)) must take exactly one route choice (\(j\)) only. So that in the combination of choices, each vehicle will only be worth \((1)\), which means that there is only one binary variable \(q\) selected.
\begin{figure}[htbp]
\centering
\includegraphics[width=0.34\textwidth, keepaspectratio]{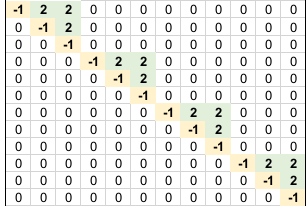}
\caption{Constraint matrix for \(n\)=4 vehicles (12 qubits)}
\label{fig 2constraint}
\end{figure}
The QUBO formula for the vehicle route selection optimization problem model is obtained from combining the cost function components (equation 5) of each segment with the constraint function (equation 7).
\begin{equation}
Obj = \sum _{s_m \in S} \left (\sum _{q_{ij} \in B_{s_m}} {w_{ij} q_{ij}} \right )^2 + K \sum _{i=1} ^{n} 
\left( \sum _{j=1} ^{3} q_{ij} -1 \right)^2
\label{eq:Obj_func}
\end{equation}
In this objective function, the parameter \(K\) is added, the Lagrange parameter or called the scaling parameter. This parameter \(K\) is used to ensure a valid priority solution on the condition that all constraints must be met (optimal objective value). In the next section, the QUBO model (Fig.3) consisting of a cost and constraint matrix will be processed by quantum annealing using access with Leap programming to the D-Wave annealer machine.

\section{Experiment \& Results}
The annealing process results will obtain the form of a binary qubit combination for the solution with the lowest energy. Each qubit combination will represent the route choice for each vehicle. Then check the value of the most optimal cost function of the route choice against the energy value generated from the annealing process. Experiment with quantum annealing for simulations on 3, 4, and 5 vehicles with the same path segment shape (12 edges). This scenario is carried out on a cost calculation model that has used segment weighting or without weighting. Finding the optimal vehicle route using quantum annealing is as follows: 
\begin{enumerate}[label=\alph*.] 
\item Identify the problem by mapping the number of vehicles, road segments, and optimization conditions (cost) to be achieved.
\item Create QUBO modeling consisting of cost and constraint functions.
\item Create a representation of the QUBO matrix used for BQM execution on the D-Wave annealer.
\item BQM execution on Leap will result in a binary combination of qubits for the solution with the lowest energy.
\item The mapping of the solution results in qubits becoming an alternative route for each vehicle.
\end{enumerate}

The next step is to compare the results of route determination using quantum annealing and manual route selection. Then the scenario is carried out on a cost calculation model that has used segment weighting or normal conditions. Cost in normal conditions (original) refers to previous research (Neukart et al., 2017), which only depends on road density. As for the use of constraints, you can use the same matrix model with the addition of an adjusted number of sizes (e.g., \(12 \times 12\) matrix for \(n=4\) vehicles and \(15 \times 15\) matrix for \(n=5\) vehicles).

\begin{figure}[htbp]
\centering
\includegraphics[width=0.45\textwidth, keepaspectratio]{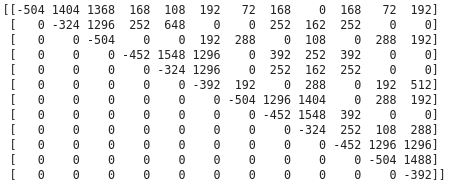}
\caption{QUBO Matrix for \(n=4\) vehicles (with weighting segments)}
\label{fig 3matrixQ}
\end{figure}
The execution of the annealing process is carried out on a D-Wave machine using 50 sampling times and \(num_reads = 50\). These results (Fig.4) get the lowest energy (E=-1184) value in the choice of vehicle route combination with the lowest cost value. The combination of route segments generated from QA processing on D-Wave provides a more optimal cost value compared to SA on Neal or with manual calculations (random).
\begin{figure}[htbp]
\centering
\includegraphics[width=0.48\textwidth, keepaspectratio]{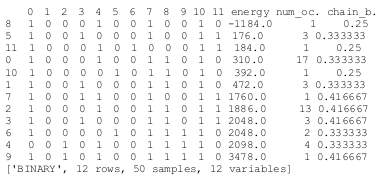}
\caption{Output annealing process for vehicle routing on D-Wave Leap at \(n=4\) vehicles (sample=50)}
\label{fig 4outputDwave}
\end{figure}

The selection of the optimal route is carried out in the case of the number of vehicles \(n=3\), \(n=4\), and \(n=5\). These results indicate the effect of alternative routes generated by considering the segment weighting parameter factors. The experiment uses a cost calculation model on the original QUBO (density) and the updated model (density, weight) with segment weighting. The cost calculation is based on the combined output of the vehicles produced using Leap (QA). In the original model, the route selection will be more accessible in the distribution of vehicles on the road segment. The choice of routes can vary and result in a better cost value. Meanwhile, the route optimization with the road segment weighting model will influence the route selection of the vehicle with a smaller weight. The road segment weight model allows overlapping vehicles to select a specific road segment, ensuring that all vehicles have exactly one route (valid result). In sampling the solution with the lowest energy, it is also possible to obtain incomplete route results for each vehicle (not valid result). So it’s necessary to post-processing the results of QA execution obtained from D-Wave Leap to be sorted based on the following conditions.
\begin{enumerate}[label=\alph*.] 
\item Check the sampling solution with the lowest energy whether there is exactly one route for each vehicle.
\item If the solution is satisfied, calculate the cost and map the solution from the qubit value to the vehicle route notation.
\item If the solution is not satisfied, look for a sampling solution with a complete route with the lowest energy requirements and the highest probability of occurrence. This solution is chosen and then mapped the qubit values into vehicle routes.
\end{enumerate}
An example of mapping the results of optimizing vehicle route selection for Fig.4 is C1=\(Q_{11}\), C2=\(Q_{22}\), C3=\(Q_{32}\), and C4=\(Q_{42}\) (with a total Cost value = 88). Quantum annealer (D-Wave Leap) execution on different sampling numbers will give different possible results but the lowest cost value. The accumulation of vehicles will begin to occur at the number of vehicles \(n = 5\). Queues of vehicles will accumulate in specific segments, such as on segments S0 and S2 (according to Fig.1). Because naturally, the route choice will prioritize the small value road weight according to the QUBO model.
\begin{figure}[htpb]
\centering
\includegraphics[width=0.43\textwidth, keepaspectratio]{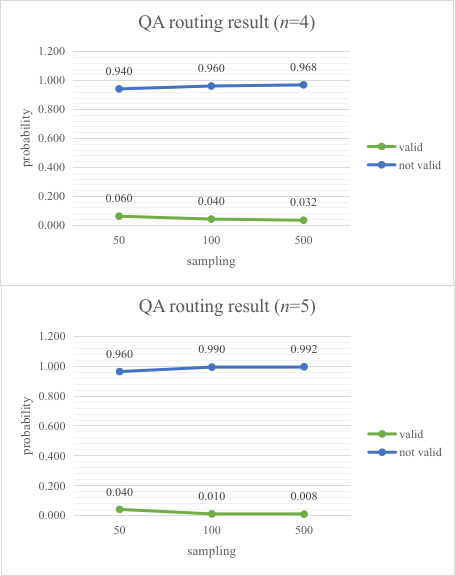}
\caption{Comparison of QA routing valid result (based on the number of sampling execution)}
\label{fig QArouting}
\end{figure}

Solving the vehicle routing problem using QA in each iteration can always provide an optimal solution. It can be seen from the comparison of using a sampling size of 50, 100, or 500 iterations (see Fig.5). However, the QUBO model in the problem requires process adjustments to reselect the solution from each iteration sample that meets valid requirements (each vehicle gets exactly one route). This valid solution will provide certainty for each route to get the least cost value. The annealing process represents the energy level produced to combine all vehicle routes with the least cost value (low road density). However, adding the weight of the road segment in the QUBO model makes it possible to have a low density in the road segment. The comparison between each iteration's cost and energy values on the D-Wave Leap shows the effect of the lowest energy generated by the optimal route with the lowest cost value. Figure 6 shows that from a series of reasonable solutions, the most optimal cost value for the choice of vehicle route is related to the lowest energy value.
\begin{figure}[htpb]
\centering
\includegraphics[width=0.41\textwidth, keepaspectratio]{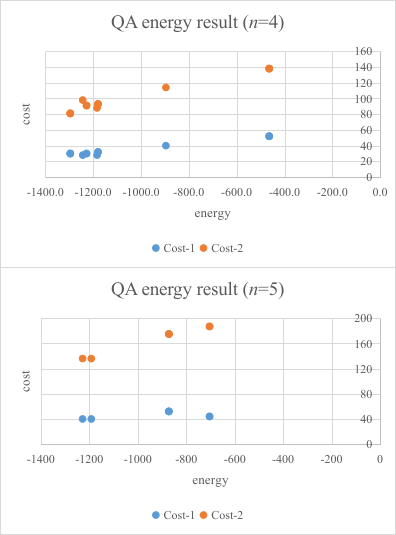}
\caption{The effect of Cost and Energy values for the optimal vehicle route solution. The result of Cost-1 is based on the density, while Cost-2 is based on the density and weight of the segment.}
\label{fig CostEnergy}
\end{figure}
Observations on the QUBO model with segment weighting will provide a different route selection solution from the original unweighted model. The accumulation of vehicles is caused by the route selection mechanism in the QUBO model, which prefers the segment with the lowest total weight. The experiments presented in Table II show that the effect of the QUBO model executed on the D-Wave quantum annealer provides a more optimal cost value than manual calculations. Compared to the manual calculation in Table II for the case of 4 vehicles, the QA cost value is about 11.11\% more optimal. While in the case of 5 vehicles, the QA cost is about 28.68\% more optimal than the manual cost calculation.
\begin{table}[htbp]
\caption{The results of vehicle route optimization using QA with road segment weighting}
\begin{center}
\begin{tabular}{c}
\includegraphics[width=0.466\textwidth, keepaspectratio]{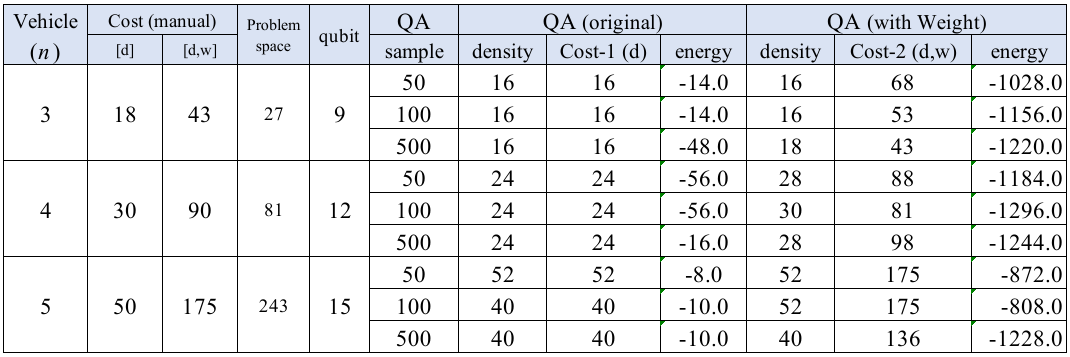}
\end{tabular}
\label{tab1}
\end{center}
\end{table}

Vehicle route mapping shows a significant increase in cost value to the number of vehicles. As Fig.1 shows, the road route only has two forks (S0 and S2), resulting in vehicle accumulation (especially in the case of n=5 vehicles). There is an average increase in road density of about 67.30\% on the route optimization results for \(n=4\) and \(n=5\) vehicles (see Table-II, density results QA with weighting). It means that about 3 to 4 vehicles simultaneously pass through the selected route for the same road segment. The proposed QUBO model only searches for the optimal segment selection based on a combination of minimal segment weights, possibly resulting in overlapping vehicle routes.

\begin{figure}[htpb]
\centering
\includegraphics[width=0.41\textwidth, keepaspectratio]{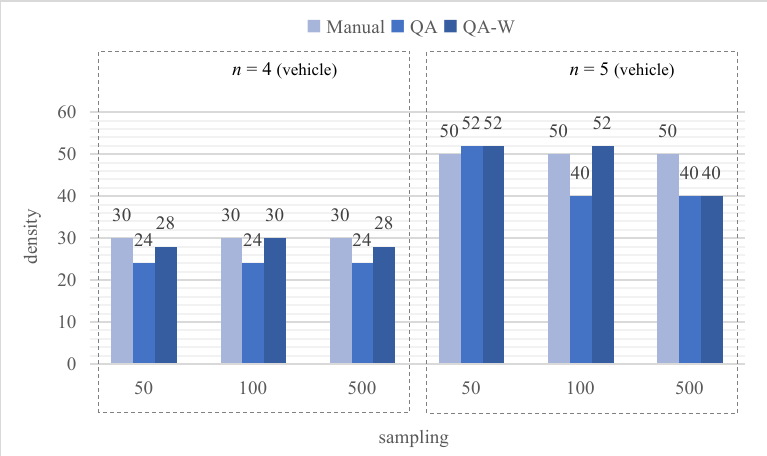}
\caption{The results of road density in experiments with 4 and 5 vehicles, show that there has been an accumulation of vehicles (in the QA results with weights) on several lanes which are influenced by the weight parameter.}
\label{fig ResultDensity}
\end{figure}

\section{Conclusions}
The research shows that QUBO can develop a vehicle route optimization model to be executed using a D-Wave quantum annealer by choosing the lowest energy as a route solution with the optimal Cost function value. The QUBO model with road segment weighting gives significant results for the route selection of each vehicle. Simulations carried out on five vehicles show that there is a possibility of increasing the density of the segment, which may cause congestion. Adding weight parameters (\(w_{ij}\)) to the road segment provides an overview of the QUBO model simulation that is as close as possible to the real conditions of searching for routes on the road. In this case, the weighting segments parameter can assume distance, road capacity, the priority of road types, and other interests. 

Future work may consider handling vehicle queues in limited segment lane conditions or overlapping vehicle routes on the same segment. The QUBO model can be proposed to handle constraints for vehicle queues, limit road capacity, traffic conditions and more varied road intersections. The current model development experiment results require \(3^n\) qubit parameters in the routing optimization solution for n vehicles. For the increasing number of vehicles, it is necessary to handle the execution of quantum annealing by hybrid processing on a D-Wave computer. The simulation in this study can be an initial reference for further development of real vehicle traffic datasets, for example, for vehicle routing on online taxi services (car or bike).

\vspace{12pt}
\printbibliography
\end{document}